\documentstyle[preprint,aps]{revtex}

\begin{document}
\title{Impurity relaxation mechanism for dynamic magnetization reversal in a single
domain grain }
\author{Vladimir L. Safonov and H. Neal Bertram}
\address{Center for Magnetic Recording Research, University of California -- San\\
Diego,\\
9500 Gilman Dr., La Jolla, CA 92093-0401, U.S.A.}
\date{\today}
\maketitle

\begin{abstract}
The interaction of coherent magnetization rotation with a system of
two-level impurities is studied. Two different, but not contradictory
mechanisms, the `slow-relaxing ion' and the `fast-relaxing ion' are utilized
to derive a system of integro-differential equations for the magnetization.
In the case that the impurity relaxation rate is much greater than the
magnetization precession frequency, these equations can be written in the
form of the Landau-Lifshitz equation with damping. Thus the damping
parameter can be directly calculated from these microscopic impurity
relaxation processes.
\end{abstract}


\section{Introduction}

The problem of magnetization reversal in a single domain ferromagnetic grain
is closely connected to the problem of relaxation. The channels of energy
transfer from the coherent magnetization rotation to the thermal bath can be
defined by interactions with itinerant electrons, spin waves (magnons),
lattice vibrations (phonons), and crystal imperfections (impurities,
defects). The theories of magnon damping in itinerant ferromagnets \cite
{reizer,solontsov} give relaxation rates $\propto k^{2}$, where $k$ is the
wave vector. Therefore this channel of energy loss is expected to be very
small for coherent magnetization rotation ($k=0$). Very large magnetization
motions are accompanied by nonlinear magnon excitations \cite{suhl,safber}.
However it has been shown that this mechanism is ineffective in extremely
small grains that undergo uniform magnetization reversal \cite{safber1}. A
mechanism for direct coupling to the lattice via magnetoelastic interaction
has been proposed \cite{suhl}.

Here we focus on the impurity relaxation processes. We believe that in fine
grains the role of imperfections is expected to be much larger than in bulk
material. In addition to paramagnetic ions (such as Fe$^{2+}$, Fe$^{4+}$, 
Mn$^{3+}$, Cr$^{2+}$, Cr$^{4+}$ etc.), 
the surface magnetic atoms are expected to behave as impurities.

It is well known that impurity magnetic ions of transition and rare-earth
elements with strong spin-orbital coupling give rise to ferromagnetic
resonance line broadening \cite{heeger,gurevich}. Two different (but not
contradictory) mechanisms of relaxation have been proposed. In the,
so-called, `slow-relaxing ion' mechanism \cite
{clogston,orbach,vanvleck,hartmann,mikhailov,saffar} small magnetization
oscillations modulate the impurity levels in the vicinity of thermal
equilibrium. This energy modulation is absorbed by the lattice at an
extremely fast rate. The `fast-relaxing ion' mechanism \cite
{degennes,white,vanvleck} is effective when the transverse relaxation rate
of the impurity is so fast that the magnetization oscillations can excite
direct transitions between impurity levels. We present a theory of
magnetization relaxation of impurities for large magnetization oscillations,
including magnetization reversal.

\section{Model}

The fundamental magnetization dynamics is described by gyromagnetic
precession:

\begin{equation}
\frac{d\,{\bf m}}{d\,t}=-\,\gamma \,{\bf m}\times {\bf H}_{{\rm eff}},
\label{LanLif}
\end{equation}
where, for coherent rotation, ${\bf H}_{{\rm eff}}=-{\partial {\cal U}}/{%
\partial {\bf M}}$ is the effective magnetic field defined by the energy
density ${\cal U}$, ${\bf M}=M_{{\rm s}}{\bf m},\quad (|{\bf m}(t)|=1)$ and $\gamma $ is the gyromagnetic ratio. We shall consider the impurities as
independent two-level systems (Fig.1). Let us write the energy density in
the form ${\cal U}={\cal U}_{{\rm m}}+{\cal U}_{{\rm imp}}$, where ${\cal U}
_{{\rm m}}$ is the magnetic energy density (including only anisotropy and
Zeeman energies). The impurities energy density can be written in the form: 
\begin{eqnarray}
{\cal U}_{{\rm imp}} &=&(\hbar /V)\sum_{j}\Omega _{0,j}n_{j}+{\cal U}_{{\rm %
ex}},  \nonumber \\
{\cal U}_{{\rm ex}} &=&\frac{\hbar }{V}\sum_{j}[{\bf w}_{j}\cdot {\bf m}%
~(n_{j}-1/2)+{\bf e}_{j}\cdot {\bf m~}\left\langle c_{j}^{+}\right\rangle +%
{\bf e}_{j}^{\ast }\cdot {\bf m~}\left\langle c_{j}\right\rangle ].
\label{imp-ham}
\end{eqnarray}
where $j$ indicates a sum over all impurities.

As illustrated in Fig.1, $V$ is the grain volume, $\Omega _{0,j}$ is the
splitting (due to crystal fields, exchange and spin-orbit interactions) of
the $j$-th impurity. ${\bf w}_{j}$, ${\bf e}_{j}$ (and its complex conjugate 
${\bf e}_{j}^{\ast }$) characterize the anisotropic exchange interaction
between each impurity (as an effective spin 1/2) and host nearest neighbor
spins (derived in the Appendix). In Eq.(\ref{imp-ham}), $n_{j}=\left\langle
c_{j}^{+}c_{j}\right\rangle $ is the upper level impurity population,
operators $c_{j}^{+}$ and $c_{j}$ describe transitions between the levels
and $\left\langle ...\right\rangle $ denotes quantum mechanical expectation
values. The contribution of the impurity interactions to the effective field
is:

\begin{equation}
{\bf H}_{{\rm eff}}=-\frac{{\partial {\cal U}}_{{\rm m}}}{{\partial {\bf M}}}%
-\frac{\hbar }{M_{{\rm s}}V}\sum_{j}\left[ {\bf w}_{j}(n_{j}-1/2)+{\bf e}%
_{j}\left\langle c_{j}^{+}\right\rangle +{\bf e}_{j}^{\ast }\left\langle
c_{j}\right\rangle \right] .  \label{effield}
\end{equation}

\section{Equations of motion}

The Eqs.(1),(3) describe the dynamic magnetization including coupling to the
impurities. The interaction of j-th impurity with the thermal bath will be
considered phenomenologically, introducing the longitudinal $\Gamma
_{\parallel ,j}$ and transverse $\Gamma _{\perp ,j}$ relaxation rates which
can depend on the impurity splitting $\Omega _{j}$, temperature $T$ and
external static magnetic field ${\bf H}_{0}$. We also assume that the
characteristic frequencies of magnetization rotation are much smaller than $%
\Omega _{j}$. Thus the kinetics of the $j$-the\ impurity population can be
defined by the following equation:

\begin{equation}
\frac{d}{dt}n_{j}=-\Gamma _{\parallel ,j}(\Omega _{j})[n_{j}-n_{T}(\Omega
_{j})].  \label{impurity}
\end{equation}
It is assumed that at each instant of time the population $n_{j}$ relaxes to
the equilibrium value $n_{T}(\Omega _{j})=[\exp (\hbar \Omega /k_{{\rm B}%
}T)+1]^{-1}$ corresponding to the dynamic splitting $\Omega _{j}=\Omega
_{0,j}+\delta \Omega _{j}(t)$, where $\delta \Omega _{j}(t)={\bf w}_{j}\cdot 
{\bf m}=w_{j,x}m_{x}(t)+w_{j,y}m_{y}(t)+w_{j,z}m_{z}(t)$. Note that the rate 
$\Gamma _{\parallel ,j}(\Omega _{j})$ also depends on this dynamic
splitting. The kinetics of ``transverse'' impurity spin components is given
by:

\begin{equation}
\left[ \frac{d}{dt}+\Gamma _{\perp ,j}(\Omega _{j})\right] \left\langle
c_{j}\right\rangle =-i\Omega _{j}\left\langle c_{j}\right\rangle -i{\bf e}%
_{j}\cdot {\bf m~[}(1-2n_{j})].  \label{kintrans}
\end{equation}

\subsection{General solution}

The general solution of Eq.(\ref{impurity}) is:

\begin{equation}
\delta n_{j}(t)=\int\limits_{-\infty }^{t}\exp [\Phi _{\parallel
,j}(t_{1})-\Phi _{\parallel ,j}(t)]~\Gamma _{\parallel ,j}[\Omega
_{j}(t_{1})]\{n_{T}[\Omega _{j}(t_{1})]-n_{T}(\Omega _{0,j})\}dt_{1},
\label{general}
\end{equation}
where

\begin{equation}
\Phi _{\parallel ,j}(t)=\int\limits_{-\infty }^{t}\Gamma _{\parallel
,j}[\Omega _{j}(t^{\prime })]dt^{\prime },\quad \delta n_{j}(t)\equiv
n_{j}(t)-n_{T}(\Omega _{0,j}).  \label{ef}
\end{equation}
Analogously, the solution of the Eq.(\ref{kintrans}) can be written as:

\begin{equation}
\left\langle c_{j}(t)\right\rangle =-i\int\limits_{-\infty }^{t}\exp [\Phi
_{\perp ,j}(t_{1})-\Phi _{\perp ,j}(t)]~{\bf e}_{j}\cdot {\bf m}%
(t_{1})~[1-2n_{T}(\Omega _{0,j})-2\delta n_{j}(t_{1})]dt_{1}
\label{gentrans}
\end{equation}
with

\begin{equation}
\Phi _{\perp ,j}(t)=\int\limits_{-\infty }^{t}\{\Gamma _{\perp ,j}[\Omega
_{j}(t^{\prime })]+i\Omega _{j}(t^{\prime })\}dt^{\prime }.  \label{efperp}
\end{equation}

By substitution of Eqs. (\ref{general}),(\ref{gentrans}) into the effective
field (\ref{effield}) and the latter into the Landau-Lifshitz equation (\ref
{LanLif}), the system of integro-differential equations for the
magnetization ${\bf m}$ is obtained.

\subsection{Slow relaxation}

In the case of small modulation $\hbar |\delta \Omega _{j}(t)|/k_{{\rm B}%
}T\ll 1$ we can write the following expansions: $n_{T}[\Omega _{0,j}+\delta
\Omega _{j}(t)]=n_{T}(\Omega _{0,j})+[\partial n_{T}(\Omega _{0,j})/\partial
\Omega _{0,j}]\delta \Omega _{j}(t)+...$ and $\Gamma _{\parallel ,j}[\Omega
_{0,j}+\delta \Omega _{j}(t)]=\Gamma _{\parallel ,j}(\Omega
_{0,j})+[\partial \Gamma _{\parallel ,j}(\Omega _{0,j})/\partial \Omega
_{0,j}]\delta \Omega _{j}(t)+...$ Utilizing only the first expansion terms,
Eq.(\ref{general}) becomes

\begin{equation}
\delta n_{j}^{(1)}(t)=\Gamma _{\parallel ,j}(\Omega _{0,j})~\frac{\partial
n_{T}(\Omega _{0,j})}{\partial \Omega _{0,j}}\int\limits_{-\infty }^{t}\exp
[-\Gamma _{\parallel ,j}(\Omega _{0,j})(t-t_{1})]~{\bf w}_{j}\cdot {\bf m}%
(t_{1})~dt_{1}.  \label{den}
\end{equation}

For the case of short memory (longitudinal relaxation $\Gamma _{\parallel
,j}(\Omega _{0,j})$ is much faster than the characteristic frequency of
magnetization rotation) we can introduce a variable $\tau =t-t_{1}$ and make
a moment expansion in powers of $\tau $. The result has the form

\begin{equation}
\delta n_{j}^{(1)}(t)=\frac{\partial n_{T}(\Omega _{0,j})}{\partial \Omega
_{0,j}}~{\bf w}_{j}\cdot \left[ {\bf m}(t)~-\frac{1}{\Gamma _{\parallel
,j}(\Omega _{0,j})}\frac{d{\bf m}(t)}{dt}+...\right]   \label{result}
\end{equation}
This first order expansion to the kinetic equation (4) for the impurity
relaxation  yields an additional term to the effective magnetic field (\ref
{effield}):

\begin{eqnarray}
\delta {\bf H}_{{\rm eff}}^{{\rm (slow)}} &\simeq &-\frac{\hbar }{M_{{\rm s}%
}V}\sum_{j}{\bf w}_{j}{\LARGE \{} [n_{T}(\Omega _{0,j})-\frac{1}{2}] +\frac{\partial
n_{T}(\Omega _{0,j})}{\partial \Omega _{0,j}}~{\bf w}_{j}\cdot {\bf m}(t)~ 
\nonumber \\
&&-\frac{\partial n_{T}(\Omega _{0,j})}{\partial \Omega _{0,j}}\frac{1}{%
\Gamma _{\parallel ,j}(\Omega _{0,j})}~{\bf w}_{j}\cdot \frac{d{\bf m}(t)}{dt%
}{\LARGE \}.}  \label{efffield}
\end{eqnarray}
The first and the second terms in Eq.(\ref{efffield}) can be included into
the effective anisotropy fields. The term containing $d{\bf m}(t)/dt$ in (%
\ref{efffield}) gives the magnetization relaxation.

\subsection{Fast relaxation}

The kinetics of the transverse impurity components (\ref{gentrans}) in the
simplest approximation can be written as:

\begin{equation}
\left\langle c_{j}(t)\right\rangle \simeq -i[1-2n_{T}(\Omega
_{0,j})]\int\limits_{-\infty }^{t}\exp [(\Gamma _{\perp ,j}+i\Omega
_{0,j})(t^{\prime }-t)]~{\bf e}_{j}\cdot {\bf m}(t_{1})~dt_{1}.
\label{trans}
\end{equation}
For a short memory the moment expansion gives:

\begin{equation}
\left\langle c_{j}(t)\right\rangle \simeq -i\frac{1-2n_{T}(\Omega _{0,j})}{%
\Gamma _{\perp ,j}+i\Omega _{0,j}}~{\bf e}_{j}\cdot \left[ {\bf m}(t)-\frac{1%
}{\Gamma _{\perp ,j}+i\Omega _{0,j}}\frac{d{\bf m}(t)}{dt}\right] .
\label{transsol}
\end{equation}
This solution (and its complex conjugate $\left\langle
c_{j}^{+}(t)\right\rangle =\left\langle c_{j}(t)\right\rangle ^{\ast }$) may
directly be added to the effective field (3).

\section{Example}

Consider for simplicity $\Omega _{0,j}=\Omega _{0}$ and the case of slow
relaxation where ${\bf w}_{j}=(w,0,0)$ and ${\bf e}_{j}=0$. Let us write the
magnetic energy in the form ${\cal U}_{{\rm m}}=K_{{\rm u}%
}(1-m_{z}{}^{2})-H_{0}M_{{\rm s}}m_{z}$, which assumes an applied field $%
H_{0}$ parallel to a uniaxial anisotropy axis. In this case the combination
of Eqs.(1),(3) and (12) may be written as:

\begin{eqnarray}
\frac{dm_{x}}{d\tau } &=&-(m_{z}+h_{0})m_{y},  \nonumber \\
\frac{dm_{y}}{d\tau } &=&(m_{z}+h_{0})m_{x}+\alpha _{1}(dm_{x}/d\tau )m_{z},
\nonumber \\
\frac{dm_{z}}{dt} &=&-\alpha _{1}(dm_{x}/d\tau )m_{y},  \label{example}
\end{eqnarray}
where $\tau =\gamma H_{{\rm K}}t,~h_{0}=H_{0}/H_{{\rm K}}$, $H_{{\rm K}}=2K_{%
{\rm u}}/M_{{\rm s}}$ and

\begin{equation}
\alpha _{1}=c_{{\rm imp}}\left( \frac{w}{\Gamma _{\parallel }}\right) \left( 
\frac{\hbar w}{k_{{\rm B}}T}\right) \frac{\exp \left( \hbar \Omega _{0}/k_{%
{\rm B}}T\right) }{\left[ \exp \left( \hbar \Omega _{0}/k_{{\rm B}}T\right)
+1\right] ^{2}},\quad c_{{\rm imp}}=\frac{\hbar \gamma N_{{\rm imp}}z_{{\rm %
imp}}}{M_{{\rm s}}V}.  \label{alpha}
\end{equation}
$N_{{\rm imp}}$ is the number of impurities in the grain and $z_{{\rm imp}}$
is the average number of magnetic neighbors for one impurity.

Fig.2 shows a typical reversal of magnetization described by the Eqs.(\ref
{example}). It is interesting to note that the Landau-Lifshitz equation with
phenomenological damping ${\cal R=-}\alpha \,\gamma \,{\bf m}\times ({\bf m}%
\times {\bf H}_{{\rm eff}})$ gives virtually the identical evolution (within
1\% of accuracy)\ of $m_{z}(\tau )$ if $\alpha =\alpha _{1}/2$. This
correspondence between $\alpha $ and $\alpha _{1}$ is exact for small
magnetization motions.

\medskip

This work was partly supported by matching funds from the Center for
Magnetic Recording Research at the University of California - San Diego and
CMRR incorporated sponsor accounts.

\appendix 

\section{Derivation of Eq.(2)}

We assume that the exchange interaction between the impurity and the
neighboring host atoms can be written in terms of an effective spin ($s=1/2$%
) of the two-state impurity ion. In this case the Hamiltonian may be written
in a general form:

\begin{equation}
{\cal H}_{{\rm ex}}=\sum\limits_{j,\nu }\sum\limits_{a_{j},a}B_{a_{j},a}(%
{\bf R}_{j},{\bf r}_{\nu })s_{a_{j}}({\bf R}_{j})S_{a}({\bf R}_{j}+{\bf r}%
_{\nu }).  \label{imp-ex}
\end{equation}
Here $B_{a_{j},a}({\bf R}_{j},{\bf r}_{\nu })$ are the parameters of
anisotropic exchange,$\ a_{j}=x_{j},y_{j},z_{j}$ are the local principal
axes for the $j$-th impurity ``spin'' and $a=x,y,z$ are the principal axes
of magnetic matrix. ${\bf s}({\bf R}_{j})=\{s_{x_{j}}({\bf R}_{j}),s_{y_{j}}(%
{\bf R}_{j}),s_{z_{j}}({\bf R}_{j})\}$ is the effective impurity spin
located at ${\bf R}_{j}$ and ${\bf S}({\bf R}_{j}+{\bf r}_{\nu })=\{S_{x}(%
{\bf R}_{j}+{\bf r}_{\nu }),S_{y}({\bf R}_{j}+{\bf r}_{\nu }),S_{z}({\bf R}%
_{j}+{\bf r}_{\nu })\}$ is the\ spin of the host matrix located in the
vicinity of $j$-th impurity.

We can express impurity spin components in terms of creation and
annihilation operators:

\begin{eqnarray}
s_{z_{j}}({\bf R}_{j}) &=&c_{j}^{+}c_{j}-1/2,\quad s_{+}({\bf R}%
_{j})=c_{j}^{+},\quad s_{-}({\bf R}_{j})=c_{j},  \nonumber \\
s_{\pm }({\bf R}_{j}) &=&s_{x_{j}}({\bf R}_{j})\pm is_{y_{j}}({\bf R}_{j})
\label{transop}
\end{eqnarray}
with the following commutation relations:

\begin{eqnarray*}
\left( c_{j}\right) ^{2} &=&\left( c_{j}^{+}\right) ^{2}=0,\quad
c_{j}c_{j}^{+}+c_{j}^{+}c_{j}=1, \\
\lbrack c_{j},c_{k}] &=&0,\quad \lbrack
c_{j},c_{k}^{+}]=(1-2c_{j}^{+}c_{j})\delta _{jk}.
\end{eqnarray*}

In the case of only coherent motion of the host spins ${\bf S}({\bf R}_{j}+%
{\bf r}_{\nu })=(V_{0}M_{{\rm s}}/\hbar \gamma ){\bf m}$, where $V_{0}$ is
the volume of one host spin. Substituting (\ref{transop}) into (\ref{imp-ex}%
) and writing ${\cal U}_{{\rm ex}}=\left\langle {\cal H}_{{\rm ex}%
}\right\rangle /V$, we obtain Eq.(\ref{imp-ham}) with $w_{j,a}=(V_{0}/\hbar
^{2}\gamma M_{s})\sum_{\nu }B_{z_{j},a}({\bf R}_{j},{\bf r}_{\nu })$, $%
e_{j,a}=(V_{0}/2\hbar ^{2}\gamma M_{s})\sum_{\nu }\left[ B_{x_{j},a}({\bf R}%
_{j},{\bf r}_{\nu })-iB_{y_{j},a}({\bf R}_{j},{\bf r}_{\nu })\right] $, $%
a=x,y,z$.

\newpage
Captions:

Fig.1

{Illustration of single domain grain and one impurity ion. }

\medskip 

Fig.2

{Field component of the magnetization versus scaled time.}

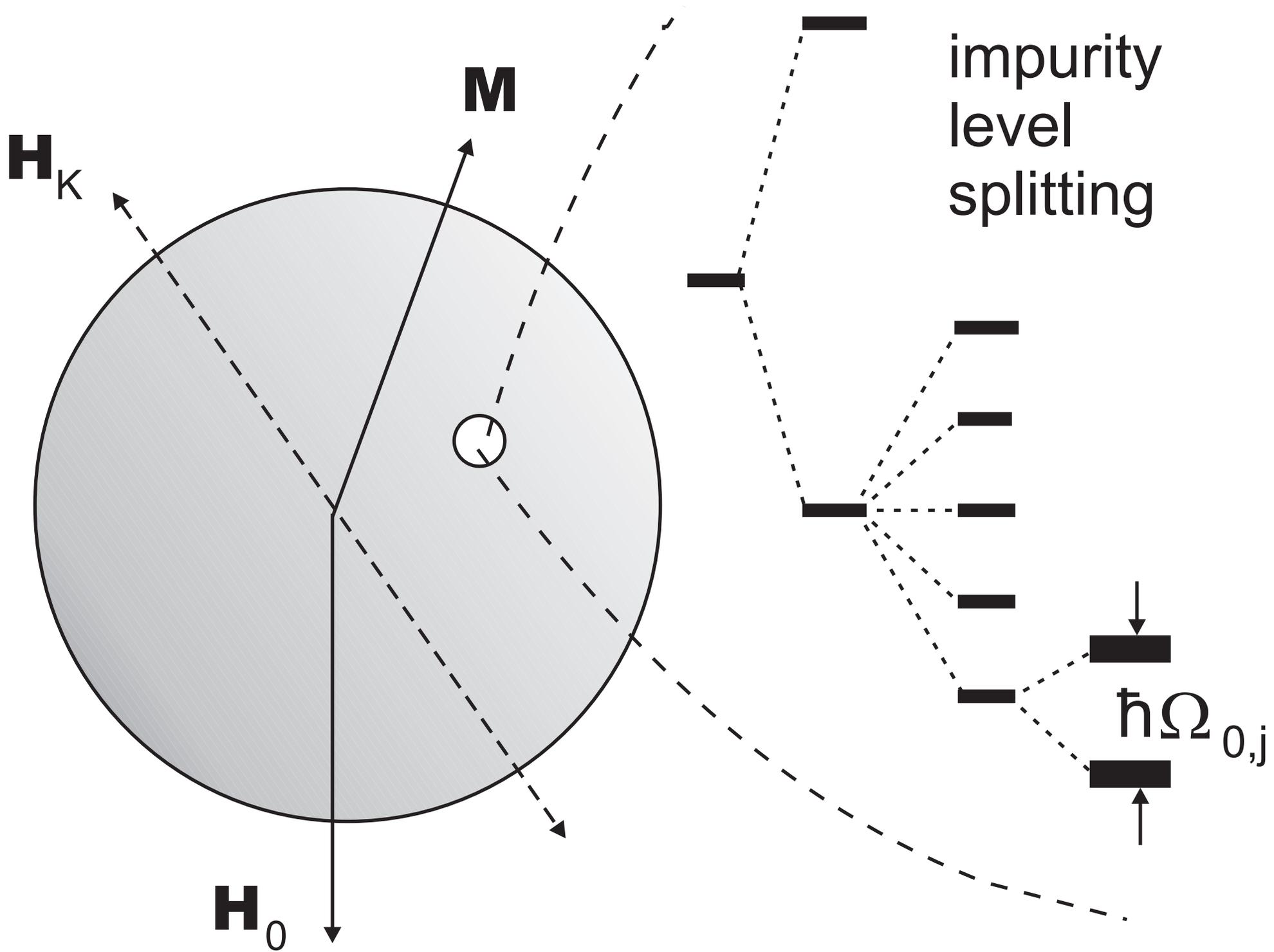
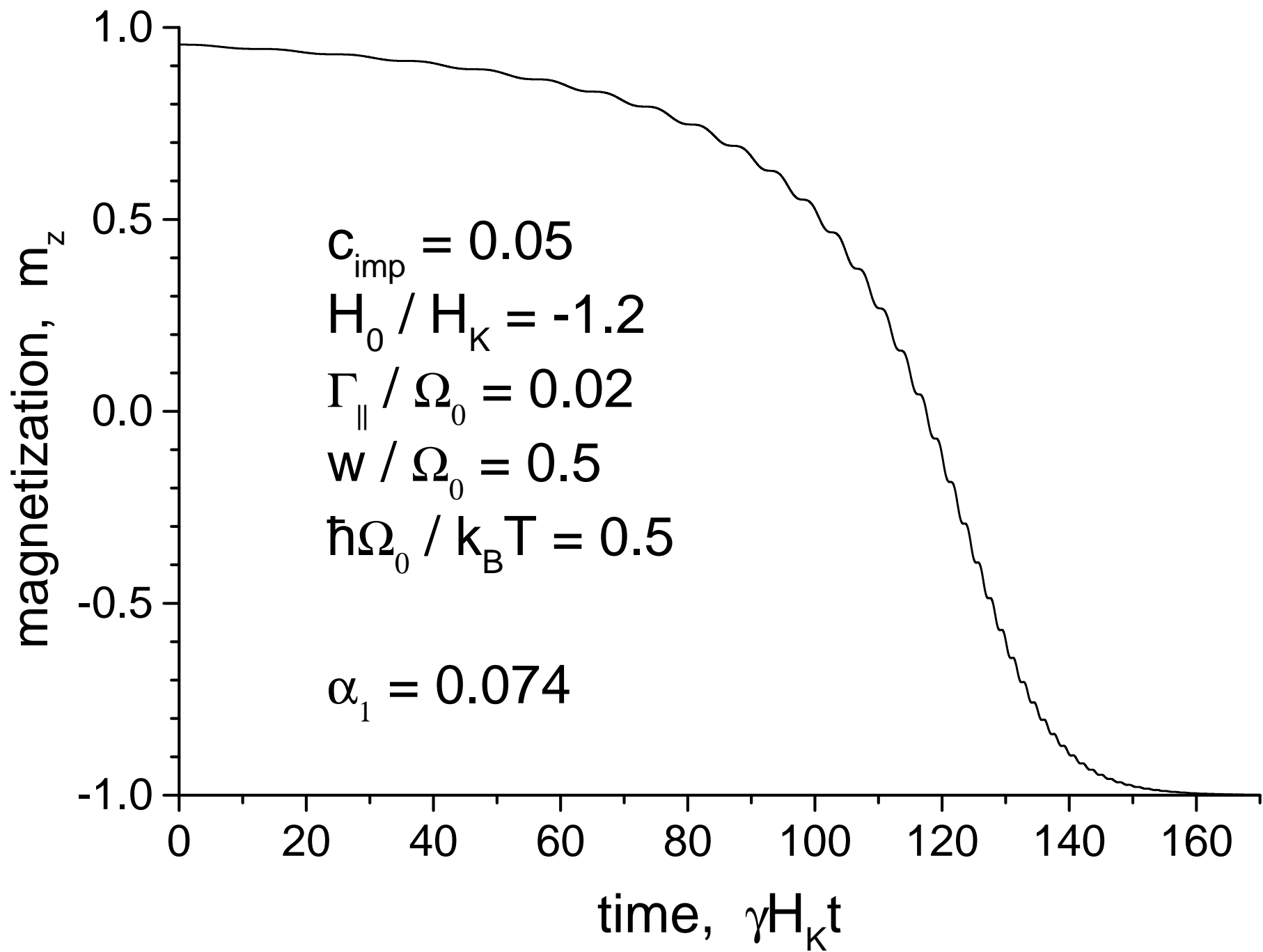
\end{document}